# Andreev Reflection Spectroscopy in Niobium Point Contacts in Magnetic Field

Y.Miyoshi, Y.Bugoslavsky, L.F.Cohen

*Blackett Laboratory, Imperial College London*

The Blonder-Tinkham-Klapwijk model of the Point-Contact Andreev Reflection spectroscopy does not provide an adequate description of experiment in the presence of magnetic field. We demonstrate this using a junction between a niobium tip and a copper film. We modify the theory to explicitly take into account the contribution to conductance that stems from the normal vortex cores in the superconductor. These results have important implications for interpretation of transport spin polarisation measurements using PCAR technique. We demonstrate that stray magnetic fields can be responsible for the experimentally observed dependence of the spin polarisation on the strength of the interface barrier, and potential misassignment of the inferred spin polarisation.

The recent theoretical [1-3] and experimental [4-18] development of point-contact Andreev reflection (PCAR) spectroscopy is driven by the interest to find materials with high degree of spin polarization $P$. Such materials can be used for integration in spin-based microelectronics [19-21]. The technique may be used to determine the spatial variation of spin-polarised current in semiconductors [22, 23]. Importantly, the technique uses a superconducting (SC) tip as a probe and either an externally applied magnetic field or the stray field from the sample may affect the behaviour of the tip. Earlier experimental works [24-28] on PCAR reported the enhancement of spectral broadening [29] with magnetic field, although without a consistent explanation of the mechanism. The influence of the magnetic field on the tip has not been addressed at all in the context of the spin polarisation measurements previously, despite the significant number of experimental studies on point-contact with ferromagnetic materials [4-16] and magnetic semiconductors [17,18].

Here we present the analysis of the behaviour of niobium tips in contact with an epitaxial copper film, in presence of an external magnetic field $H$, applied parallel or perpendicular to the tip. To describe the field dependence of PCAR spectra, we introduce an extension to the Blonder-Tinkham-Klapwijk (BTK) model [24] that takes into account the effect of normal vortex cores in the SC tip explicitly. We show that the effect of vortices has profound implications for the analysis of spin polarisation in ferromagnetic samples.

Point contacts were created by driving a mechanically sharpened Nb tip (made of a 0.25mm-diameter, 99.9% pure Nb wire), into the surface of a 60-nm thick, (100)-oriented copper film. The conductance vs voltage ($G(V)$) dependences were recorded using a standard lock-in technique, at T = 4.2K.

Point-contact spectra with the field applied parallel or perpendicular to the tip (Fig. 1), show gradual suppression of the spectral features with increasing magnetic field. All spectra are normalized by the background conductance taken at approximately 20mV, where the conductance is independent of voltage. The position of the peaks stays constant at approximately ±2mV up to 0.4T and starts to narrow progressively at higher fields. Although spectra at $H$ higher than 1.4T are not shown, the peaks are still identifiable at 1.8T. As the peaks start to converge, the spectral feature becomes smaller, and disappears at 2 T, indicating the local upper critical field, $H_{c2}$ [31].

As in our previous work [28], we assume that the contact area (~10 μm in diameter) contains multiple randomly-distributed individual junctions. Together they probe a large area of the vortex lattice compared to the inter-vortex separation, yielding an overall spectrum, which is effectively an average over the vortex lattice. In order to take into account the normal cores of vortices, we consider the total conductance $G(V)$ to be a sum of the normal and SC channels ($G_N$ and $G_S(V)$, respectively). $G_N$ is invariant with voltage, representing an ohmic normal-state junction. We further assume that $G_S(V)$ is described by the BTK model with certain values of the order parameter $\Delta$, interface barrier $Z$ and smearing parameter $\omega$. We treat the spectral broadening by calculating a convolution between the zero-temperature SC density of states (DOS) and a Gaussian function of width $\omega$ [28]. This generic method accounts for all sources of broadening. We note that the thermal part ($k_B T = 0.36$ meV; $k_B$ is the Boltzmann constant) constitutes only about a half of the observed total zero-field $\omega$. We assume that the non-thermal contribution is mainly due to the interface scattering.

Since the density of vortices $n = \frac{B}{\Phi_0} \approx \frac{H}{\Phi_0}$ (where $B$ is magnetic induction and $\Phi_0$ is the flux quantum), we can assume that the contribution of the normal channel is proportional to $H$. At $H_{c2}$ the SC channel should disappear. Therefore, $G(V)$ can be written as

$$G(V) = hG_N + (1-h)G_S(V), \qquad (1)$$

where $h=H/H_{c2}$. At $H=0$ and $H=H_{c2}$, the conductance is BTK and ohmic, respectively, and in between there is a progressive suppression of the Andreev-reflection features with $H$.

Fig.1 presents the experimental data obtained with two orthogonal field orientations (parallel and perpendicular to the tip) and the corresponding best fits to Eq.1. The fits were obtained by fixing the value of $h$ in accordance with the applied field and varying the parameters $\Delta$, $Z$ and $\omega$.

The quality of fit in itself can not be taken as a proof of the validity of the two-channel model, because fits of similar quality can be obtained using the conventional formulae that neglect the effect of vortex cores. The difference between the two models becomes apparent when the field dependences of the inferred parameters are analysed.

In Fig 2 we plot the fitting parameters for the series of curves of Fig 1(a). Taken at face value, the results of the conventional BTK fits could be interpreted as apparent field dependences of the junction parameters, $\Delta_a$, $Z_a$ and $\omega_a$ (open symbols in Fig. 2). All these dependences are strong, but most importantly, the interface barrier, $Z_a$ appears to diverge as the field approaches $H_{c2}$. Such behaviour is unphysical, as the interface and the mechanical properties of the junction are not affected by the field.

However, when the SC part of the conductance is first extracted from the experimental data according to Eq. (1), and then fitted to the BTK model, the field dependences of the parameters change considerably (solid symbols in Fig 2). $Z$ is now independent of field, reinforcing the validity of the two-channel model. The effective order parameter attains approximately parabolic dependence on $H$ and the increase of the experimental spectral broadening ω becomes less pronounced, although still present. These results demonstrate the significance of including the influence of the normal vortex cores for the analysis of the field dependences of the superconductor properties inferred from the in-field PCAR spectroscopy.

In the case of perpendicular field, although the zero-field junction properties (primarily the barrier $Z$) were different, the overall evolution of the spectra, and the apparent field dependences of the junction parameters were similar to those shown in Fig. 2. This similarity suggests that the point-contact transport is not directionally selective, and therefore it couples efficiently to the vortex cores irrespectively of the mutual orientation of the tip and the vortices.

Earlier work by Naidyuk et al [26] on the type I superconductor Zn demonstrates that no spectral broadening in field occurs when there is a sharp first order transition to the normal state. The disparity in the behaviour of Type I and Type II superconductors convincingly demonstrates that it is the introduction of vortices that causes the strong field dependence of the PCAR spectra.

The fact that the BTK formalism does not apply to a superconductor in the mixed state can be explained as follows. In the presence of vortices, the order parameter $\Delta$ of the superconductor (and, consequently, the electron DOS) varies spatially. Golubov and Kupriyanov [32] have calculated numerically the DOS at different distances from the vortex centre, and obtained the DOS averaged over the unit cell of the vortex lattice. The density at Fermi level, $N(0)$, increases with field following a closely linear dependence [33]. The complete treatment of Andreev reflection in magnetic field would require using the appropriate theoretical DOS in calculating the BTK-like expressions for reflection and transmission coefficients at the interface. We show however that to a good approximation this computationally intensive problem can be circumvented by using the much simpler two-channel model.

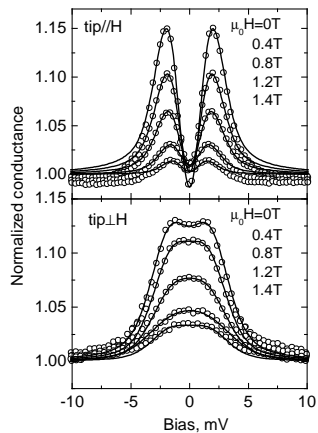

Figure 1. Experimental data for a Nb/Cu point contact, in field applied parallel (a) and perpendicular (b) to the Nb tip. The solid lines are best fits to Eq.1.

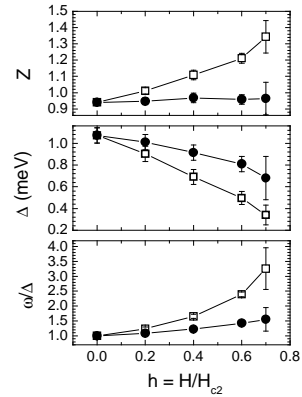

Figure 2. Fitting parameters (barrier $Z$, SC order parameter $\Delta$ and relative spectral broadening, $\omega/\Delta$) obtained from the experimental data of Fig 1(a) using the BTK model and neglecting the effect of the magnetic field (open symbols), and the more adequate two-channel model (solid symbols). Quantitatively similar results were obtained in field applied perpendicular to the Nb tip.

Shown in Fig 3 (a) is a one-dimensional pictorial representation of the actual variation of the order parameter (solid line) and the simplified rectangular profile used implicitly in the two-channel model (dotted line). In our model the continuous variation of the order parameter is replaced by perfectly normal vortex cores situated in the otherwise homogeneous superconductor. In Fig 3 (b) we show the comparison between the exact theoretical DOS calculated in [32] and the function that results from the two-channel model. Our model proves to describe correctly the principal features of the mixed-state DOS, namely the sub-gap excitations (*i.e.*, the contribution of the cores) and the effective broadening of the DOS features in increasing field. In contrast, the conventional BTK model fails to describe the low-energy part of the DOS.

We note that in principle it is formally possible to reproduce the vortex-lattice average DOS, and hence the experimental results, assuming a very strong and field-dependent life-time broadening in the SC [25]. However there is a difficulty explaining the origin of this effect [24], whereas our interpretation based on the vortex contribution is clear and consistent.

We can therefore conclude that the dominant effect in the field dependence of the PCAR spectra is due to the normal-state excitations in the vortex cores, and the two-channel model reflects the physical essence of the phenomenon.

We can now discuss the implications of the magnetic field (either applied externally or the stray field of a magnetic sample) for the measurements of spin polarisation. The stray field may be particularly important. Spin-polarised materials are usually ferromagnets with large spontaneous magnetisation. The stray field is negligible if the film is uniformly magnetised in the plane, but it is high in thin films with out-of-plane magnetisation, such as FePd [34]. In samples with an in-plane easy axis the stray field may be high locally in the vicinity of domain walls. Also, the tip can be subjected to a strong local field if it creates an indentation in the surface of the film fully magnetised in the plane. The exact details of the field distribution are strongly dependent on the material and sample geometry. Here we present a generic analysis of the potential impact the magnetic field can have on the analysis of PCAR data on spin polarised materials

The BTK model has been extended to include the transport spin polarisation $P$ by Mazin et al [1]. We use their expressions to generate the superconducting term in Eq (1) (now with non-zero $P$), and produce a series of spectra for varying $h$. As before, we address the dominant effect, and assume for simplicity that $\Delta =1.5$ meV, $Z=0.5$, $\omega =0.5$ meV and $P$ are field-independent. The generated curves are fitted to the zero-field generalised BTK model [1] in order to highlight the apparent effects due to neglecting the magnetic field. The fitting procedure yields for each curve a set of apparent parameters, $\Delta_a$, $Z_a$, $P_a$ and $\omega_a$ that are dependent on $h$. The apparent value of $\Delta_a$ decreases only slowly with field to reach ~ 1.3 at h=0.5, and $\omega_a$ does not show any strong filed dependence either. However $P_a$ and $Z_a$ are generally strongly dependent on the field. Reporting $P(Z)$ dependence is a usual way experimental PCAR results are presented for magnetic samples [10]; therefore in Fig. 4 we plot $P_a$ versus $Z_a$ with $h$ as an implicit parameter.

There are several important observations. For low $P_0$ we recover the approximately parabolic dependence, similar to what has been reported experimentally (e.g., for nickel in [7]). If such dependence is inferred from a number of measurements taken at different locations on the sample surface with different arrangement of domain walls in the junction area, the effective field on the tip will change between the measurements, causing the apparent $Z$ and $P$ to change. At higher $P_0$ (~0.6) the apparent $P(Z)$ dependence flattens out, and at $P$ close to 1 the gradient of the $P_a(Z_a)$ dependence becomes positive. We can conclude that whereas the magnetic mechanism can be a possible candidate to explain the observations in the relatively low-$P$ materials, it is unlikely to apply in the case of highly spin-polarised $CrO_2$ [7]. Contrary to the common assumption, if the magnetic mechanism is responsible for the $P(Z)$ dependence, the extrapolation to $Z=0$ can be misleading, especially so for low-$P$ materials.

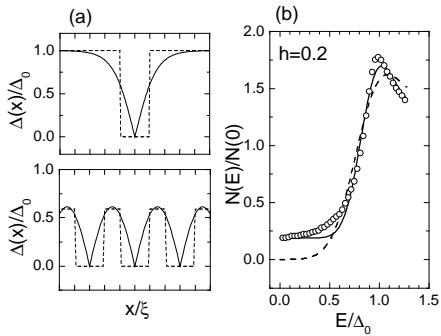

Figure 3. (a) Schematic spatial variation of the normalized order parameter in the presence of Abrikosov vortices is shown by the solid line (top: low field; bottom: high field). In the two-channel model this is approximated the rectangular function shown as the dashed line. $\xi$ is the coherence length of the superconductor, and $\Delta_0$ is the zero-field order parameter. (b) Theoretical DOS in the mixed state of a superconductor at $h$=0.2 (Ref [32]; circles). The two-channel model approximates the exact DOS with the function shown as the solid line. The conventional BTK model (dashed line) cannot reproduce the sub-gap excitations associated with the vortex cores.

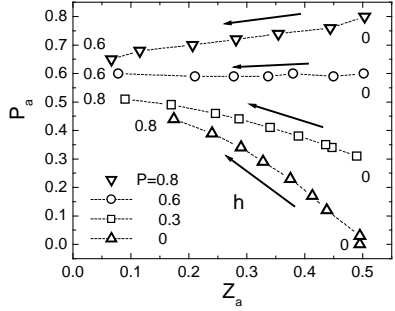

Figure 4. Variation of the apparent BTK fitting parameters $P_a$ and $Z_a$ for the simulated data with fixed $P_0$ and $Z_0$ and varying reduced field $h$. $Z_0 = 0.5$ for all data and $P_0$ is 0.8, 0.6, 0.3 and 0 (top to bottom). The arrows indicate increasing magnetic field, $h=H/H_{c2}$, which is changed in steps of 0.1. The numbers by the data points show the values of $h$. At small $P_0$, as $h$ increases $Z_a$ and $P_a$ vary to produce a smooth nearly parabolic dependence. The variation of $P_a$ becomes weak for data with large starting $P_0$.

In summary, we have studied the behaviour of the point-contact spectra of Nb tip/Cu film as a function of applied magnetic field, at 4.2K. The key effect in the variation of the spectra with field is due to the normal conduction channel provided by the cores of Abrikosov vortices. We present a simple extension to the BTK model, which adequately describes the experiment. Ignoring the vortices in the analysis may lead to serious artefacts, especially when the spin polarisation is measured in a ferromagnetic material with high saturation magnetisation.

We are grateful to I.I.Mazin for his suggestion to analyse the effect of magnetic field on the inferred spin polarisation, and to A.A.Golubov and B.E.Nadgorny for helpful discussion. We thank C.H.Marrows for supplying us with the copper film for this study. This work has been supported by the UK Engineering and Physical Sciences Research Council.

### References

[1] I. I. Mazin, A. A. Golubov, and B. Nadgorny, J. Appl. Phys. 89, 7526 (2001)
[2] I. Zutic and O. T. Valls, Physical Review B 61, 1555 (2000).
[3] K. Xia, P. J. Kelly, G. E. W. Bauer, and I. Turek, Phys Rev Lett 89, 166603 (2002).
[4] R. J. Soulen, J.M. Byers, M.S. Osofsky, B. Nadgorny, T. Ambrose, S.F.Cheng, P.R. Broussard, C.T. Tanaka, J. Nowak, J.S. Moodera, A.Barry, J.M.D.Coey, Science 282, 85 (1998).
[5] S. K. Upadhyay, A. Palanisami, R. N. Louie, and R. A. Buhrman, Phys. Rev. Lett. 81, 3247 (1998).
[6] B. Nadgorny, R.J.Soulen, M.S.Osofsky, I.I.Mazin, G.Laprade, R.J.M.van de Veerdonk, A.A.Smits, S.F.Cheng, E.F.Skelton, S.B.Qadri. Phys. Rev. B 61, R3788 (2000).
[7] Y. Ji, G.J.Strijkers, F.Y.Yang, C.L.Chien, J.M.Byers, A.Anguelouch, G.Xiao, A.Gupta, Phys. Rev. Lett. 86, 5585 (2001).
[8] B. Nadgorny, I.I. Mazin, M.Osofsky, R.J.Soulen, P.Broussard, R.M.Stroud, D.J.Singh, V.G.Harris, A.Arsenov, Y.Mukovskii. Phys. Rev. B 63, 184433 (2001).
[9] G.J.Strijkers, Y.Ji, F.Y.Yang, C.L.Chien, J.M.Byers, Phys. Rev. B 63, 104510 (2001).
[10] C. H. Kant, O.Kurnosikov, A.T.Filip, P.LeClair, H.J.M.Swagten, W.J.M.de Jonge, Phys. Rev. B 66, 212403 (2002).
[11] W.R. Branford, S.K.Clowes, Y.V.Bugoslavsky, Y.Miyoshi, L.F.Cohen, A.V.Berenov, J.L.MacManus-Driscoll, J.Rager, S.B.Roy, J. Appl. Phys. 94, 4714 (2003).
[12] B. Nadgorny, M.S. Osofsky, D.J. Singh, G.T.Woods, R.J.Soulen, M.K.Lee, S.D.Bu, C.B.Eom. Appl. Phys. Lett. 82, 427 (2003).
[13] L. Ritchie, G.Xiao, Y.Ji, T.Y.Chen, C.L.Chien, M.Zhang, J.L.Chen, Z.H.Liu, G.H.Wu, X.X.Zhang. Phys. Rev. B 68, 104430 (2003).
[14] P. Raychaudhuri, A. P. Mackenzie, J. W. Reiner, and M. R. Beasley, Phys. Rev. B 67, 020411 (2003).
[15] N. Auth, G. Jakob, T. Block, and C. Felser, Phys. Rev. B 68, 024403 (2003).
[16] S. K. Clowes, Y.Miyoshi, Y.Bugoslavsky, W.R.Branford, C.Grigorescu, S.A.Manea, O.Monnereau, L.F.Cohen. Phys. Rev. B. 69, 214425 (2004).
[17] J. G. Braden, J.S.Parker, P.Xiong, S.H.Chun, N.Samarth. Phys. Rev. Lett. 91, 056602 (2003).
[18] R. P. Panguluri et al., cond-matt 0403451, (2004).
[19] G. A. Prinz, Science 282, 1660 (1998).
[20] J. de Boeck and G. Borghs, Physics World 12, 27 (1999).
[21] S. A. Wolf, D.D.Awschalom, R.A.Buhrman J.M.Daughton, S. von Molnar, M.L.Roukes, A.Y.Chtchelkanova, D.M.Treger, Science 294, 1488 (2001).
[22] J. Seufert, G.Bacher, H.Schomig, A.Forchel, L.Hansen, G.Schmidt, L.W.Molenkamp. Phys. Rev. B 69, 035311 (2004).
[23] J. Sinova, D. Culcer, Q. Niu, et al. Phys Rev Lett 92, 126603 (2004)
[24] Yu.G.Naidyuk, R.Haussler, H.v.Lohneysen. Physica B 218, 122 (1996)
[25] Y. de Wilde, T.M.Klapwijk, A.G.M.Jansen, J.Heil, P.Wyder. Physica B 218, 165 (1996)
[26] Yu.G.Naidyuk, H.v.Lohneysen, I.K.Yanson. Phys Rev B54, 16077 (1996)
[27] R.S. Gonnelli, D. Daghero, A.Calzolari, G.A.Ummarino, V. Dellarocca, V.A.Stepanov, J.Jun, S.M.Kazakov, J.Karpinski J.Phys Rev B69, 100504 (2004)
[28] Y.Bugolsavsky, Y.Miyoshi, G.K.Perkins, A.D.Caplin, L.F.Cohen, A.V.Pogrebnyakov , X.X.Xi. Phys Rev B 69, 132508 (2004)
[29] R.C.Dynes, V.Narayanamurti, J.P.Garno, Phys. Rev.Lett. 41, 1509 (1978).
[30] G. E. Blonder, M. Tinkham, and T. M. Klapwijk, Phys. Rev. B 25, 4515 (1982).
[31] The local value of $H_{c2}$ of the junction is approximately 2.2 times higher than the bulk $H_{c2}$ of the wire material, as obtained from the heat capacity measurements. We attribute the enhancement of $H_{c2}$ to strong plastic deformations of the tip due to the pressure necessary to establish a stable point contact.
[32] A. A. Golubov and M. Yu. Kupriyanov, J. Low Temp. Phys. 70 83 (1988).
[33] A.A.Golubov, private communication
[34] C.H. Marrows, B.C. Dalton. Phys. Rev. Lett. 92 097206 (2004)